\documentclass[aps,prb,twocolumn,groupedaddress,epsf]{revtex4}

\usepackage{graphicx}
\usepackage{dcolumn}
\usepackage{bm}

\begin{document}

\title{Effect of parallel magnetic field on the Zero Differential Resistance State}
\author{N. Romero, S. McHugh, M. P. Sarachik and S. A. Vitkalov }
\address{Physics Department, City College of the City
University of New York, New York, New York 10031}
\author{A. A. Bykov}
\address{Institute of Semiconductor Physics, 630090 Novosibirsk, Russia}

\date{\today }

\begin{abstract}

The non-linear zero-differential resistance state (ZDRS) that occurs for highly mobile two-dimensional electron systems in response to a dc bias in the presence of a strong magnetic field applied perpendicular to the electron plane is suppressed and disappears gradually as the magnetic field is tilted away from the perpendicular at fixed filling factor $\nu$. Good agreement is found with a model that considers the effect of the Zeeman splitting of Landau levels enhanced by the in-plane component of the magnetic field.

\end{abstract}

\maketitle


The nonlinear properties of highly mobile electrons in
two-dimensional AlGaAs/GaAs heterojunctions have been the focus 
of a great deal of recent attention. Strong oscillations of
the longitudinal resistance induced by microwave radiation
have been found\cite{Zudov, Ye} at magnetic fields satisfying the condition
$w=nw_c$, where $w$ is the microwave frequency and $w_c$ is the
cyclotron frequency ($n =1,2,...$). At high levels
of microwave excitation the minima of the oscillations can
reach a value close to zero.\cite{Mani, Zudov2, Dorozhkin, Willett} This so-called zero
resistance state (ZRS) has stimulated extensive theoretical
attention.\cite{Andreev, Durst, lei, Anderson, Shi, Vavilov1, Dmitriev} 

Interesting nonlinear phenomena have also been found
in response to a dc electric field.\cite{Yang, Bykov, Zhang, Chiang}
Oscillations of the longitudinal resistance, periodic as a function of the
inverse magnetic field,
have been observed at relatively high dc bias satisfying the
condition $n\hbar w_c= 2R_cE_H$; here $R_c$ is the cyclotron radius
of electrons at the Fermi level and $E_H$ is the Hall electric
field induced by the dc bias in the magnetic field. This
effect has been attributed to horizontal Landau-Zener
tunneling between Landau levels, tilted by the Hall electric
field $E_H$.\cite{Yang} 
Another notable nonlinear effect is a strong
reduction of the longitudinal resistance by considerably
smaller dc electric fields. \cite{Bykov, Zhang, Chiang} This effect has been
attributed \cite{Zhang} to spectral diffusion of electrons in a dc electric field.\cite{Dmitriev}
Electron spectral diffusion occurs in the presence of a strong magnetic field where the density of states (DOS) oscillates due to Landau quantization. The oscillations in the DOS result in an oscillatory structure of the non-equilibrium electron distribution function. When a dc electric field $E_{dc}$ is applied, electrons diffuse from low energy regions (occupied levels) to high energy regions (empty levels) through elastic scattering between electrons and impurities. Inelastic scattering limits this process, forcing the electron distribution function back to thermal equilibrium. 
This effect also accounts for a nonlinear electron state with zero differential resistance (ZDRS) 
which has been recently identified.\cite{vitkalov1,zudov2008}
The ZDRS exhibits strong dependences on both temperature and magnetic field \cite{vitkalov1} through the strong dependence of the electron spectral diffusion on these parameters.\cite{Dmitriev,Zhang,Vavilov2} 
In this paper we study the effect of an in-plane magnetic field on the ZDRS and the nonlinearity of the 2D electron system induced by a dc bias. This research was also motivated by apparent differences between reported measurements of the ZRS induced by microwave radiation in response to an in-plane magnetic field.\cite{duH,maniH} 

The sample used in this experiment was cleaved from a wafer of a high-mobility GaAs quantum well grown by molecular beam epitaxy on a semi-insulating (001) GaAs substrate.  The quantum well was $13$ nm wide, the electron density n$=9.2\times 10^{15}$ m$^{-2}$, and the mobility $\mu$=85 m$^2$/Vs at T$=1.7$ K. Measurements were carried out at T$=1.7$ K in magnetic
fields up to $9$ T on $50  \mu$m-wide Hall bars with a distance of 250 $\mu$m between potential contacts. The differential longitudinal resistance was measured at a frequency of 77 Hz in the linear regime.  Direct electric current (dc bias) was applied simultaneously with an ac excitation through the same current leads (see inset to Fig. 1(a)).

\begin{figure}
\includegraphics[width=0.50\textwidth]{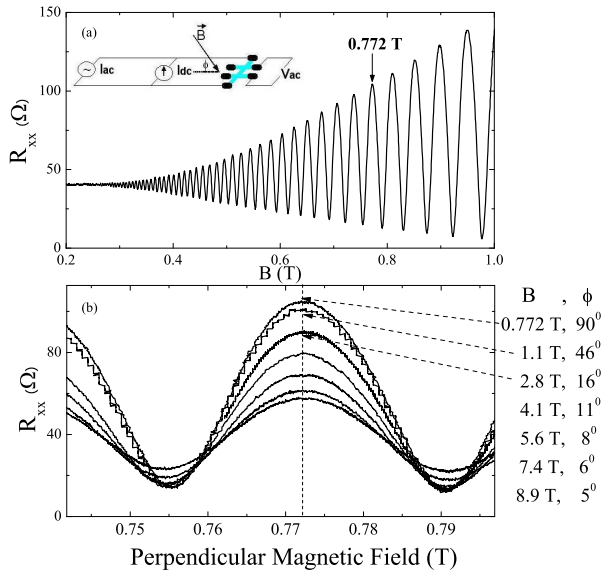}
\caption{ \label{fig1}
(a) Quantum oscillations of the resistance at T$=1.7$ K for magnetic field applied perpendicular to the electron ($\phi=$90$^o$); the arrow denotes the field of the Shubnikov-de Haas maximum for which subsequent data were obtained (see text); the inset is a schematic of the experimental set-up; (b) Resistance $R_{xx}$ plotted as a function of the perpendicular component $B_{\perp}$ of the magnetic field  for magnetic field applied at various angles $\phi$ with respect to the electron plane.  The legend lists the angle $\phi$ and the $total$ magnetic field at the maxima (marked by the dashed line). Data were taken at T$=1.7$ K with zero dc bias.
}
\end{figure}

\begin{figure}
\includegraphics[width=0.45\textwidth]{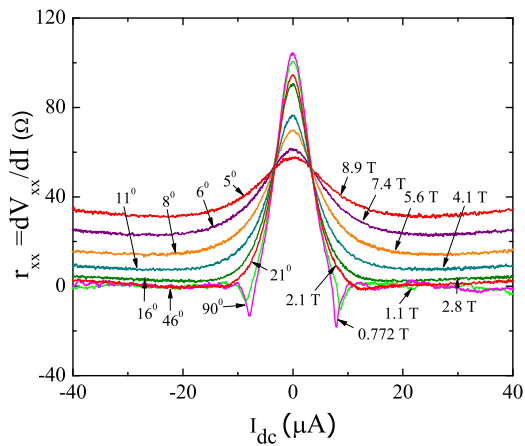}
\caption{ \label{fig2}  
Differential resistance versus dc bias for different angles $\phi$ between the magnetic field and the 2D electron plane, where the perpendicular magnetic field $B_{\perp}=0.772$ T (marked by an arrow in Fig. 1(a)) corresponds to a SdH maximum and is fixed for all curves. The temperature T$=1.7$ K.
}
\end{figure}

Figure 1(a) shows quantum oscillations of the resistance at T$=1.7$ K as a function of magnetic field applied perpendicular to the electron plane ($\phi=$90$^o$). The arrow denotes the Shubnikov-deHaas maximum at $B_{\perp} = 0.772$ T at which the measurements reported below were taken.
In Fig. 1(b) the resistance $R_{xx}$ is plotted as a function of the perpendicular component $B_{\perp}$ for magnetic field applied at different angles with respect to the plane.  While all curves display a maximum at $B_{\perp}=0.772$ T, as expected, the magnitude of the resistance peaks at $0.772$ T decreases as the angle $\phi$ decreases from $90^o$ and the total magnetic field increases. For the measurements reported below, we rotated the sample and simultaneously varied the magnitude of the total magnetic field in order to fix the perpendicular magnetic field component at $0.772$ T while changing the in-plane magnetic field.  The filling factor $\nu$ is thus fixed for all curves in Fig. 1(b), while the Zeeman splitting $\Delta _{Z}=g \mu _B B$ is different for different curves due to its dependence on the total magnetic field $B$. 
 
Figure 2 shows the differential resistance $r_{xx}=dV_{xx}/dI$ as a function of $dc$ bias at $T=1.7$ K for different angles $\phi$ and fixed perpendicular magnetic field $B_{\bot}=0.772$ T corresponding to the Shubnikov de Haas (SdH) oscillation maximum indicated by the arrow in Fig. 1(a).  Note that the total magnetic field (denoted on the right-hand side of Fig. 2) and its in-plane component both increase as the angle $\phi$ decreases.
The differential resistance $r_{xx}$ initially decreases with increasing bias $I_{dc}$ for all angles.  For a perpendicular magnetic field ($\phi=90^o$) $r_{xx}$ exhibits a reproducible negative spike at $I_{dc}=7.9 \mu$A and then stabilizes near zero. This is the zero differential resistance state. As the angle $\phi$ between the magnetic field and the plane is decreased, the spike gradually disappears and is no longer observable at  $\phi=21^{o}$. For smaller angles the differential resistance $r_{xx}$ is increasingly positive as the angle $\phi$ decreases, and a shallow minimum develops at large bias.
 
It is interesting to compare our results for the zero differential resistance state (ZDRS) with those reported in Ref. [21] and Ref. [22] for the effect of in-plane field on the zero resistance (ZRS) state. Both experiments were performed in magnetic fields smaller than those used in our experiments.  Mani \cite{maniH} tilted the sample at an angle $\theta$ with respect to the magnet axis and microwave propagation direction. The ZRS was observable with the oscillatory pattern unchanged at a tilt angle of $\theta=80^o$ ($\phi=90^o-\theta=10^o$), and vanished only at $\theta\approx90^o$. The disappearance of the ZRS at $\theta\approx90^o$ was attributed to the vanishing of the photon flux through the two dimensional electron system rather than to the in-plane magnetic field. Yang et al. \cite{duH} employed a two axis system to provide perpendicular and parallel field components. They report the gradual reduction of the microwave induced ZRS and its disappearance when a parallel magnetic field ($B_{||}\approx0.5$T) is applied.   Our results are qualitatively similar to those of Yang et al. \cite{duH}: we find that the ZDRS decreases and disappears gradually with increasing in-plane magnetic field component, while Mani \cite{maniH} reported quenching of the ZRS only at $\theta\approx90^o$.  It is possible, however, that stronger magnetic fields, comparable to those applied in our experiments, are required to reduce the dc induced nonlinearity for $\theta<90^o$.

We suggest that the suppression of the nonlinear response of the system and the disappearance of the zero differential resistance at small angles $\phi$ are due to the change of the bias-stimulated spectral diffusion of the electrons \cite{Dmitriev,Zhang,vitkalov1} caused by the increase of the in-plane magnetic field component. We consider Zeeman splitting of the Landau levels as the main mechanism leading to a decrease of the variations of the spectral diffusion with energy and, thus, to the reduction of the nonlinearity. Below we compare numerical simulations of the spectral diffusion with experiment. Good agreement is found. 

To estimate quantitatively the effect of Zeeman splitting on the spectral diffusion we begin by analyzing the change in the electron spectrum induced by the Zeeman effect. As the angle of the applied magnetic field is tilted away from the perpendicular and the total magnetic field is increased, the oscillations of the density of states (DOS), $\nu(\epsilon)$, split into spin up and spin down components, as seen in Eq. 1. This leads to a reduction in the modulation of the DOS amplitude.\cite{vitk2001}
In order to calculate the DOS we use a gaussian approximation given by\cite{raich}

\begin{eqnarray}
  \tilde{\nu}(\epsilon)=\frac{\nu (\epsilon)}{\nu _0} & = & \frac{\sqrt{\omega _{c} \tau _q}}{2}\bigg(exp\left(-\frac{(\epsilon/\hbar+ \Delta _{z}/\hbar-n\omega _c)^2 }{\omega _{c}/\pi \tau _q} \right) \nonumber \\
  & & {} + exp\left(-\frac{(\epsilon/\hbar- \Delta _{z}/\hbar-n\omega _c)^2}{\omega _{c}/\pi \tau _q} \right)\bigg), 
\end{eqnarray} 
where $\tilde{\nu}(\epsilon)$ is the dimensionless DOS normalized by the value of the DOS at zero magnetic field, $n$ is an integer, $\tau_q$ is the quantum or single particle relaxation time and $\omega_c$ is the cyclotron frequency. The parameter  $\hbar/ \tau_q$ determines the width of the Landau levels and is obtained from comparison with experiment (shown below). A similar value of $\hbar/\tau_q$ is also found by comparison of the experiment with the self consistent Born approximation of the density of states.\cite{ando} 
 
The inset to Fig. 3 shows the results of numerical simulation of the effect of  Zeeman splitting on the density of electron states (DOS) in our sample. It can be seen that the modulation of the DOS is weaker for smaller angles ($\phi=8^o$) corresponding to stronger Zeeman spin splitting. 
Figure 3 presents the angular dependence of the maximum value of the differential resistance $r_{xx}=dV_{xx}/dI$ at $B_{\perp}=0.772$ T obtained from the curves presented in Fig. 1(b). At fixed filling factor the resistance decreases with decreasing angle $\phi$ (with a consequent increase of the total magnetic field applied). Based on the evolution of the density of states displayed in the inset, the theoretically expected values of resistance are denoted by the circles of Fig. 3 for comparison. The resistance was estimated using a simplified expression for the longitudinal conductivity in strong magnetic fields $(\omega_c \tau_p \gg 1)$: \cite{Dmitriev}
$$
    \sigma_{xx}= A \times \int \sigma(\epsilon)(-d f/d\epsilon) d\epsilon, \eqno{(2)} 
$$
where $\sigma (\epsilon )= \sigma_D \tilde{\nu}(\epsilon)^2$, $\sigma_D=e^2 \nu _{0} v_F^2/2\omega_c^2 \tau_{tr}$ is the Drude conductivity in a perpendicular magnetic field $B_{\perp}$. The free parameter $A$ accounts for possible memory effects \cite{Vavilov2} and other deviations from Drude behavior in the presence of strong magnetic fields.\cite{Shklov} The parameters $A$ and $\tau_q$ (the quantum scattering time), were chosen to provide a good fit between experiment and theory for the angular dependence of the resistance at $B_\perp=0.772$ T.  From the comparison above, we obtain the electron $g$-factor, $g=-0.475$, which is very close to values obtained in other experiments.\cite{stormer} We find good agreement between experiment and theory (see Fig. 3). Thus, we are able to attribute the decrease of the SdH maxima with increasing in-plane magnetic field component to the Zeeman effect.

\begin{figure}
\includegraphics[width=0.45\textwidth]{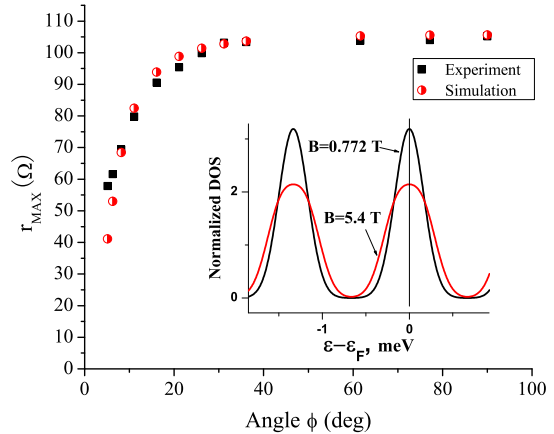}
\caption{ \label{fig3}  
The differential resistance $r_{MAX}$ of the  
quantum oscillation maximum at $B_\perp=0.772$ T plotted as a function of the angle $\phi$ between the total magnetic field and the electron plane. The squares are the experimental results and the circles represent the numerical simulation. The inset shows the density of electron states (normalized to its value at zero magnetic field) in a fixed perpendicular magnetic field  $B_\perp=0.772$ T and different total magnetic fields, as labeled.}
\end{figure}
 
In order to estimate how the electron spectral diffusion and the nonlinearity of the 2D electron system in the presence of a magnetic field is affected by the Zeeman effect, we solve numerically the spectral diffusion equation for the electron distribution function $f(\epsilon)$: \cite{Dmitriev} 
$$
-\frac{\partial f(\epsilon)}{\partial t}+E_{dc}^2\frac{\sigma _{dc}^D}{\nu _0 \tilde{\nu}(\epsilon)}\partial _{\epsilon}\left[\tilde{\nu} ^2(\epsilon) \partial _{\epsilon}f(\epsilon)\right]=\frac{f(\epsilon)-f_T(\epsilon)}{\tau_{in}},
\eqno{(3)}
$$
where $f_T(\epsilon)$ is the Fermi distribution and $E_{dc}$ is the bias-induced electric field. For the normalized DOS, $\tilde{\nu}(\epsilon)$, we use the DOS obtained above from a comparison with the linear response (see Eq. 1, Eq. 2 and Fig. 3). Spectral diffusion is a result of elastic scattering between electrons and impurities in the presence of a bias-induced electric field $E_{dc}$; it is limited by inelastic processes, which force the distribution function back to thermal equilibrium. We use the inelastic relaxation time $\tau_{in}$ as a fitting parameter. The solution of the diffusion equation $f(\epsilon)$ at $t \gg \tau_{in}$ is then inserted into Eq. 2 in order to obtain the resistivity at different dc biases. 

\begin{figure}
\includegraphics[width=0.45\textwidth]{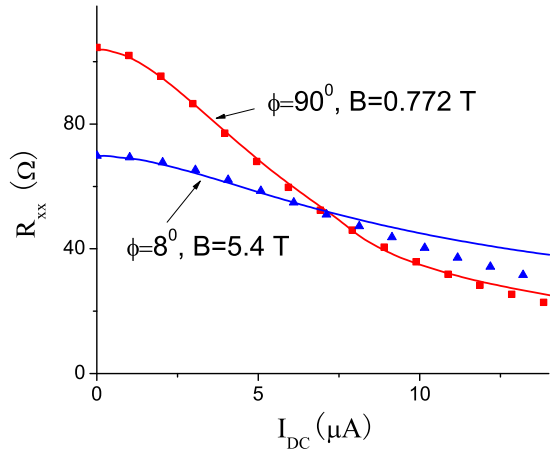}
\caption{ \label{fig4}  
The solid curves show the measured resistance $R_{xx}=V_{xx}/I_{dc}$ as a function of dc bias in fixed perpendicular magnetic field for two different total magnetic fields (angles $\phi$), as labeled; T$=1.7$ K. The symbols denote the numerical solution of the spectral diffusion equation; $\tau_q=5.1$ ps; $\tau_{in}=2.6$ ns at $\phi=90^o$ and $\tau_{in}=2.7$ ns at $\phi=8^o$.}
\end{figure}

Figure 4 shows experimental and numerical results for the longitudinal resistance $R_{xx}$=$V_{xx}/I$ as a function of the dc bias plotted for two different angles, $\phi=90^{o}$ and  $\phi=8^{o}$. The vertical scale is fixed by the comparison with the linear response (by the choice of the two parameters $A$ an $\tau_q$ in Eq. 1 and Eq. 2). The horizontal scale is chosen to provide the best fit between theory and experiment. The best result is obtained for the inelastic time $\tau _{in}=2.6\times 10^{-9}$ s at $\phi=90^{o}$ and for $\tau _{in}=2.7\times 10^{-9}$ s at $\phi=8^{o}$. There is good agreement between theory and experiment at small dc bias. At higher dc bias deviations become evident that are larger for smaller angles $\phi$. We suggest that these deviations are due to additional nonlinear mechanisms that occur at higher dc bias \cite{Vavilov2,Dmitriev2} which have not been treated in this paper.

In conclusion, the effect of a dc electric field on the longitudinal resistance of a highly mobile two-dimensional (2D) electron system in GaAs quantum wells was studied.  We observe a zero-differential resistance state in response to a direct current when the magnetic field is perpendicular to the electron plane. At fixed filling factor the nonlinearity of the 2D electron system decreases and the zero differential resistance state disappears gradually as the total magnetic field is increased and tilted toward the 2D plane.  Numerical simulations of the spectral diffusion in the presence of Zeeman splitting of the DOS in a high magnetic field provide a good fit to the experimental observations.

We thank Jing-Qiao Zhang for his help with the numerical simulations.  
This work was supported by NSF grant DMR 0349049, U. S. DoE grantDOE-FG02-84-ER45153 and RFBR project No.06-02-16869.


\end{document}